# Gel-state nucleation in multilamellar vesicles of dimyristoylphosphatidylcholine and its relation to the critical temperature: A lattice model and microcalorimetry

**Dmitry P. Kharakoz, [1]  Maria S. Panchelyuga, [1] Elizaveta I. Tiktopulo [2] and Elena A. Shlyapnikova [1]**

[1] Institute of Theoretical and Experimental Biophysics of RAS
[2] Institute of Protein Research of RAS, Pushchino, Russia

## Abstract

Differential microcalorimetric measurements have been performed in aqueous dispersions of dimyristoylphosphatidylcholine (DMPC) multilamellar vesicles in a wide range of temperatures covering the whole transition between the lamellar gel and lamellar fluid states (the chain-melting/ordering transition). The shape of calorimetric curves was analyzed in a temperature range some distance away but close to the chain-ordering transition point where small nuclei of gel (solid state) are formed. In this range, where the fraction of the "new-state" lipid is small, the nucleation process can be considered independent of the interlayer interactions and determined mainly by the lateral interactions. The small-scale nucleation was analyzed in terms of a two-dimensional Ising-like lattice model. The gel-fluid contact energy related to the critical temperature (in terms of Ising model) was estimated for DMPC studied here and dipalmitoylphosphatidylcholine studied earlier. The contact energy was found to be not high enough to provide the discontinuous, first-order character of transition. Therefore, the signs of first-order character observed in the vesicles made of these lipids are not due to lateral but rather interlayer 3D-interactions. The extrapolation to longer lipids shows that the transition discontinuity inherent to a lipid layer (i.e., determined by the lateral interactions themselves) is expected for the saturated phosphatidylcholines having more than 16-18 C-atoms per chain. Interestingly, it is the saturated chains of this length that are the most abundant in biological membranes. Probably, the biological membrane "prefers" to be near the critical state where the system is the most responsive to physical actions.

# 1. Introduction

The chain-melting/ordering transition in lipid membranes evidently plays an important role in cell physiology [1−3]. To understand physical grounds, which underlie its physiological function, we need to study the factors determining the physiologically important features of the transition [2, 3]. Among these features is its high cooperativity: the transition occurs with a jump-like change of enthalpy and volume as if it is of first-order [4−6]. However, the transition is not absolutely discontinuous. Moreover, extensive fluctuations of area and volume are observed in the liquid-crystalline state (*l*) close to the point of transition to gel (*g*) [2−9]. These properties are typical of second-order transitions. From these reasons, the transition is often considered as a pseudocritical phenomenon [7−10]. The question of whether the transition is discontinuous depends on the position of the critical point—does the transition occur below or above the critical temperature? Unfortunately, this problem cannot be solved by means of mere phenomenological considerations, directly from the experimentally determined temperature dependence of relevant parameters. The direct phenomenological analysis of the whole transition curves is aggravated by the possible existence of long-lived defects in the real lipid systems (which, for example, may be due to impurities); they can obscure the transition discontinuity making it apparently diffuse [11]. Another aggravating factor is that in typical classes of biologically relevant lipids, such as diacylphosphatidylcholines (PCs), an intermonolayer interaction of steric nature takes place. This interaction results from the interplay between two structural requirements: on the one hand, the lateral packing mismatch between bulky polar heads and thin nonpolar chains leads to a tendency of the monolayer to bend upon solidification; on the other hand, the bilayer and, hence, the comprising monolayers must have a common local curvature. A compromising structure can represent a spatially correlated modulation of the lateral packing order (briefly reviewed in refs. [12, 13]). It causes an essentially three-dimensional intermediate structure to appear—the ripple phase. The chain-ordering thus splits onto two steps called main transition (between the liquid-crystalline and ripple states) and pre-transition (between the ripple and gel states).

Therefore, studying the mechanism of the transition cooperativity, one should distinguish between the contributions made by lateral interactions and interlayer coupling into the determination of the transition type. Our work is aimed at estimating the energy of lateral, in-plane interactions, influencing the critical temperature, and its dependence on the length of lipid chains in saturated PCs. Our approach is based on the assumption that some distance away from the main transition temperature (where nuclei of the solid state are much less in size than the period of ripples, and their mole fraction is low), the intermonolayer coupling of steric nature can be neglected. Then, a simple lattice model can be used as the theoretical basis for interpretation. Such an approach has been earlier suggested and approved for lipid bilayers in a series of works [14−16]; this approach is based on the Monte Carlo simulation in an Ising-like two-state two-dimensional model.

In this work, we studied dimyristoylphosphatidylcholine (DMPC) multilamellar vesicles in aqueous dispersions by means of adiabatic scanning calorimetry. The calorimetric data on this lipid system and those earlier reported on dipalmitoylphosphatidylcholine (DPPC) [17] were analyzed within the framework of above-mentioned theoretical model. From comparison of the experimental and theoretical data, we estimated the energy of the lateral nearest-neighbor interaction in the two lipid systems and came to the following conclusions. The main transition in both lipids occurs above the critical temperature inherent to individual lipid layers. In other words, the lateral interaction itself cannot cause a discontinuous (first-order) transition. The extrapolation to longer PCs shows, however, that the first-order character may be an intrinsic property of the monolayer made of lipids having at least 17 C-atoms per chain or more. An analogous relation to the chain length is observed in the phenomenon of anomalous bending fluctuations: they vanish when the length reaches 18 C-atoms [18]. Interestingly, the saturated chains with this critical length are the most abundant in biological membranes. Probably, the biological membrane "prefers" to be near the critical state where the system is the most responsive to physical actions.

## 2. Experimental

### *2.1. Preparation of lipid vesicles*

Synthetic DMPC was purchased from Avanti and used without further purification. The lipid powder was dried over 1 week under $P_2O_5$ in a vacuumed desiccator prior to use. Multilamellar vesicles were prepared by mixing the powder with double-distilled water, incubating the mixture for 30 min at 30°C (6° above the chain-melting point), and vortexing it periodically. The lipid concentration was determined by weight to an accuracy of 1% (molecular mass is 678.1). Prior to be placed into a measuring cell, the sample was degassed in a gastight syringe to reduce the probability of the bubble formation.

### *2.2. Calorimetric measurements*

Heat capacity was measured with a differential scanning calorimeter Scal-1 (Scal Co. Ltd., Pushchino, Russia) having glass capillary cells with 0.3 ml sample volume. The capillaries were horizontally oriented (a precaution against the non-uniform sedimentation of lipid dispersions along the capillary). Baseline reproducibility of the device, $\sim 0.3 \times 10^{-6}$ W, corresponded to the relative error in heat absorption 0.01% relative to absorption in water, at the scan rate 0.125 K/min. The response time of the device was 30 s; this time delay might cause broadening of the heat absorption peak by a value of the order of 0.06 K at our scan rate. This distortion was not essential even for the sharpest part of the curve, while the deviations in the range of our interest were about 5 K wide (see Results).

An extra constant pressure of 2 atm was applied over the liquids in the cells. The molar heat capacity of lipid was calculated from the measured heat absorption by means of the equation

$$C_P = \Delta/m + C_{P,w} \rho_w V,$$

where $\Delta$ is the entire heat capacity difference between the cell filled with a lipid sample and that filled with water; $m$ is the lipid mass in the sample cell; $\rho_w$ and $C_{P,w}$ are the density and specific heat capacity of water, respectively; $V$ is the lipid specific volume [19]. Lipid mass in the active part of calorimetric cell was considered independent of temperature, as the lipid sediment did not flow out of the cell upon expansion of water at our conditions. The data on density and heat capacity of water were taken from Ref. [20]. Lipid volume was assumed 0.95 cm$^3$/g at 30 ºC [21], and its temperature dependence was estimated assuming proportionality between the volume and enthalpy [7, 22]: $V = 0.000758\,H$, where enthalpy $H$ is an integral of $C_P$ over temperature. Heat capacity $C_P$ was thus obtained by means of a convergent iteration algorithm in which $V = V_{30ºC} + 0.000758(H - H_{30ºC})$. The calculated volume agreed with experimental data [21, 22, 23].

## 3. Model

The model of lipid membrane described in this subsection is similar to that used earlier [14, 24]. A lipid monolayer is modeled by a flat triangular lattice with $N$ lattice points ($N$ changed from 100×100 up to 800×800, depending on the proximity to the transition point where the coherence length increases). Each point is occupied by a lipid chain. Every chain can exist in two states corresponding to gel (*g*) or liquid crystal (*l*). Periodic boundary conditions are used to eliminate the effect of edges. The expansion work is not considered. The states of monolayers in a bilayer, as well as the states of bilayers in a multilamellar system are considered to be noncorrelated. Therefore, the full energy ($E$) of the system consists of two terms:

$$E = -\Delta s\,(T - T_0)\,N_l + E_{gl} N_{gl} \tag{1}$$

The first term is responsible for the intrinsic contribution of all lattice points; $N_l$ is the number of the points in state $l$; $T$ and $T_0$ are the current and transition temperatures, respectively; $\Delta s$ is a proportionality coefficient whose meaning is the entropy of $g$-$l$ transition of a single lipid chain (the value $\Delta s$ can be taken from the experimental data). The second term is responsible for the contribution of $N_{gl}$ unlike nearest-neighbor contacts into $E$; $E_{gl}$ is the energy of such a contact. Written in reduced terms, the full energy is

$$E^* = \Delta E^* N_l + \varepsilon N_{gl}, \qquad (2)$$

where

$$E^* = E / k_B T,$$

$$\varepsilon = E_{gl} / k_B T,$$

and

$$\Delta E^* = -\Delta s (T - T_0) / k_B T \qquad (3)$$

($k_B$ is Boltzmann constant). Therefore, $-\Delta E^*$ has a meaning of a reduced temperature, and $\varepsilon$ is the reduced $g$-$l$ contact energy. The latter parameter is a measure of the energetic disadvantage of unlike contacts and, hence, it determines the extent of transition cooperativity [14]. This model is formally equivalent to the Ising model in a temperature-dependent external field. For studying lipid membranes, thus modified Ising model has been first suggested by Doniach [24].

Thermodynamic equilibrium in the system was achieved by means of Monte Carlo simulation (MC) in the canonical ensemble using Metropolis algorithm [25]. One MC cycle consisted of $N$ trials so that each of $N$ lattice points was probed to be changed through one cycle. To randomize the trials, the sequence of the probed points was randomly selected in each cycle. The simulations were performed with two variable parameters: $\varepsilon$ and $\Delta E^*$. The parameter $\varepsilon$ varied within 0.41-0.57 covering the critical value $\varepsilon_{crit} = 0.549$ above which the transition becomes discontinuous (see below). For each $\varepsilon$, simulation started at a minimum value $\Delta E^* = -0.7$ corresponding to the maximum temperature in the set of our experimental data. After the sufficiently long trajectory of simulation, $\Delta E^*$ was increased to the next value, thus changing stepwise from $-0.7$ towards zero; this is equivalent to cooling the system down. The cooling run was stopped at $\Delta E^* = 0$ (or earlier if the behavior of the system became bistable, indicating that the coherence length became comparable with the lattice dimensions). Afterwards, a heating run started through the same set of $\Delta E^*$ but in inverse direction, to check the quality of equilibration and reversibility of the system. For each $\Delta E^*$, the two simulation trajectories were compared in order to exclude from further consideration their initial parts where the fraction of gel-state points differed in the two runs. The remaining parts were used for calculating the mean equilibrium fraction of lattice points in $g$-state, $X$. No less than $10^3$ configurations were used for getting the mean.

Note that the term $E_{gl}$ has the meaning of extra energy per an unlike contact. This makes a difference from the term commonly used to define the interaction in Ising models, $J$, the term meaning that the energy of a contact changes from $-J$ to $+J$ upon the transition from the like to unlike case; hence, $E_{gl} = 2J$. It should also be noted that there is a critical value of the contact energy parameter, $\varepsilon = \varepsilon_c$, when the transition temperature $T_0$ gets equal to the critical one, $T_c$; therefore, the transition becomes discontinuous at $\varepsilon > \varepsilon_c$. The value $\varepsilon_c$ follows from a strong theoretical solution given by Fisher for Ising model in triangle lattice [26]: $J/k_B T_c = 0.2746$. Considering the system at $T_0$ and using the above-defined reduced terms, one comes to $T_c = 0.5 E_{gl} /0.2746 k_B = \varepsilon T_0/0.549$, or

$$T_c = T_0 \varepsilon/\varepsilon_c, \qquad (4)$$

where $\varepsilon_c = 0.549$.

# 4. Results

## *4.1. Calorimetric data and the parameters of direct g-to-l transition*

Calorimetric curves are presented in Fig. 1. The shape of the curves is similar to that reported earlier [27], the difference being within the experimental error reported by the authors. The two peaks of heat capacity observed in the figure correspond to two transitions: between the gel and ripple states called pre-transition (at 13.5 °C) and between the ripple and liquid-crystalline states called main transition (at 23.9 °C). Our purpose is to analyze the shape of the pretransional region above the main transition point, where small clusters of the solid-like nuclei form. Following Ref. [17], we assume the structure of the solid-state nuclei embedded in the parent liquid-crystalline phase to resemble the gel rather than the ripple state (this is reasonable because the nuclei are much smaller than the size of ripples). Therefore, of our interest is to determine the thermodynamic functions for a hypothetical *direct* transition from an "undisturbed" gel to an "undisturbed" liquid crystal, bypassing intermediate states, like ripple phase or pretransitional structures. We define the "undisturbed" states as those where the temperature dependences of heat capacity are, within the error, linear functions, as shown with dashed lines in Fig. 1A. As seen from this figure, the so-defined "undisturbed" gel and liquid crystal exist at temperatures below 9 °C and above 35 °C, respectively.

It should be noted that the slope of $C_P(T)$ for the gel state is determined with the accuracy of about 20% because of a short range used for approximation. Our examination shows, however, that the curvature of $H(T)$ for the gel state is affected by this error to an extent, which is not essential for further consideration. Fig. 1B presents the experimental curve for $H(T)$ (solid line) and the curves corresponding to the hypothetical undisturbed states (dashed lines). These functions are obtained by integration: $H(T) = \int_{T_{ref}}^{T} C_P \, dT$, where $C_P$ stands for either real system or its hypothetically "undisturbed" states; $T_{ref}$ is an arbitrary reference temperature. Gibbs free energy change ($\Delta G$) upon the direct gel-to-liquid crystal transition was calculated as $\Delta G(T) = \int_{T_{ref}}^{T} C_P \, dT - T \int_{T_{ref}}^{T} \frac{\Delta C_P}{T} \, dT$, where $\Delta C_P$ is the heat capacity change upon the transition [19]. We have found that within a wide range (10-35 °C) covering main transition and pretransitional phenomena, $\Delta G$ is an almost linear function of temperature:

$$\Delta G \approx -\Delta S \, (T - T_0), \qquad (5)$$

where $T_0$ and $\Delta S$ are the direct g-l transition temperature and molar entropy, respectively. The midpoint of the direct transition (where $\Delta G=0$) was found to be 1.0 degree below the main transition temperature. Numerical data are colleted in Table 1.

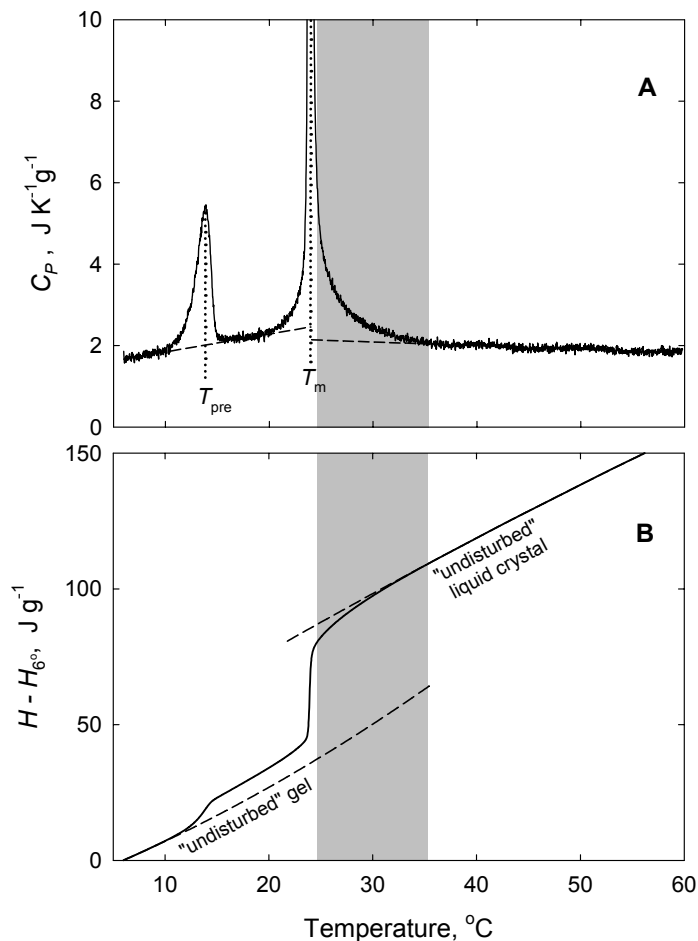

Fig. 1. Specific heat capacity (A) and enthalpy (B) as functions of temperature. The temperatures of main ($T_m$) and pre-transition ($T_{pre}$) are defined as half-transition points; the temperature of the hypothetical direct gel-to-liquid-crystalline state transition ($T_0$) is determined as described in Results. Dashed lines represent the extrapolated temperature dependences for hypothetical "undisturbed" states. The range of detectable pretransitional change caused by small-scale nuclei of gel is indicated in gray halftone.

Pronounced deviations of experimental curves from the lines of "undisturbed" states are observed near the transition point in a wide range of temperatures, much wider than the range of the transition itself (Fig.1). The deviations taking place above the transition point and caused by small-size nuclei of the gel state are the matter of our further consideration. Following the earlier suggestion [17], we consider these deviations in terms of an *apparent* mole fraction of gel-state, $X_{app}$, determined from enthalpy as follows:

$$X_{app} = (H_{\text{"undisturbed" liquid crystal}} - H_{\text{observed}}) / \Delta H \qquad (6)$$

where $\Delta H = (H_{\text{"undisturbed" liquid crystal}} - H_{\text{"undisturbed" gel}})$ is the g-to-l transition enthalpy. The results are discussed below.

### *4.2. Monte Carlo simulation data*

Fig. 2 presents a typical curve for the actual mole fraction of gel, $X(-\Delta E^*)$ obtained through MC simulation (dashed line). For the purpose of comparison with experimental data, we should

Table 1. Characteristics of the direct transition from "undisturbed" gel to "undisturbed" liquid crystal in DMPC and DPPC membranes.

| quantity | value [a] | | unit |
|---|---|---|---|
| | DMPC | DPPC | |
| Direct transition temperature, $T_0 - 273.15$ | $22.9 \pm 0.1$ [b] | $40.1$ [c] | °C |
| (Main transition temperature, $T_m - 273.15$ [d]) | $(23.9 \pm 0.1)$ | $(41.2$ [c]$)$ | °C |
| Free energy, $\Delta G$ | $-0.111 \times (T - T_0)$ | $-0.142 \times (T - T_0)$ [c] | kJ mol$^{-1}$ |
| Entropy, $\Delta S$ | $0.111 \pm 0.002$ [b] | $-0.142$ [c] | kJ K$^{-1}$mol$^{-1}$ |
| Enthalpy, $\Delta H = T_0 \Delta S$ | $32.9 \pm 0.4$ [b] | $44.6$ [c] | kJ mol$^{-1}$ |
| Reduced contact energy at $T_0$, $\varepsilon$ [e] | $0.495 \pm 0.005$ | $0.53 \pm 0.005$ | |
| Critical temperature, $T_c - 273$ [f] | $-7 \pm 3$ | $29 \pm 3$ | °C |
| Critical chain length where $T_c = T_0 \approx T_m$ | 16.5-18 [g] | | number of C-atoms |

*a.* The errors correspond to the scatter of the results of independent experimental runs. *b.* Determined from linear approximation of $\Delta G(T)$; Eq. 5. *c.* From ref. [17]. *d.* Main transition temperature is presented for the purpose of comparison; defined as the half-transition point. *e.* Estimated by visual examination of Fig. 3. *f.* Calculated from $\varepsilon$ according to Eq. 4. *g.* Linear extrapolation of $\varepsilon$ to longer chains.

transform the actual mole fractions to the apparent ones in the same manner as we treated the experimental data. Namely, the system in a range far from the transition region should be considered "undisturbed", implying that each of the two states contains a small fraction of lattice points in the opposite state. When analyzing calorimetric data, we considered the system to be undisturbed in the temperature ranges approximately 15 K away from $T_0$; i.e., $|T - T_0| > 15$ K. The model system should be considered "undisturbed" in the same range, which in our case corresponds to $|\Delta E^*|>0.30$ (calculated with Eq. 3 in which $\Delta s$ is substituted with $0.5\Delta S/N_A$, where $\Delta S$ is the experimental molar entropy taken from Table 1 and $N_A$ is Avogadro constant). The function $X(-\Delta E^*)$ in these ranges is approximated by second polynomials (dotted lines in Fig. 2), similarly to how the experimental enthalpy was approximated. The difference between the two approximating lines is considered to be the complete transition effect expressed in terms of actual mole fractions, $\Delta X$ (a vertical arrow in Fig. 2). The apparent mole fraction of gel state is then defined as

$$X_{app} = (X_{\text{"undisturbed" liquid crystal}} - X_{observed}) / \Delta X \qquad (7)$$

where $X_{observed}$ is the actual fraction of $g$-state lattice points observed in simulation, $X_{\text{"undisturbed" fluid}}$ is that extrapolated for "undisturbed" $l$-state. Resulting function $X_{app}(-\Delta E^*)$ is drawn in Fig. 2 with solid line.

## 5. Discussion

A family of curves $X_{app}(-\Delta E^*)$ for different $\varepsilon$ is presented in Fig. 3. With $\varepsilon$ increasing, the transition is getting sharper and becomes even discontinuous when $\varepsilon > 0.55$. This result is in agreement with the known critical value $\varepsilon_c = 0.549$ (see Model). What is the value of $\varepsilon$ that corresponds to the experimental data: is it below or above $\varepsilon_c$?

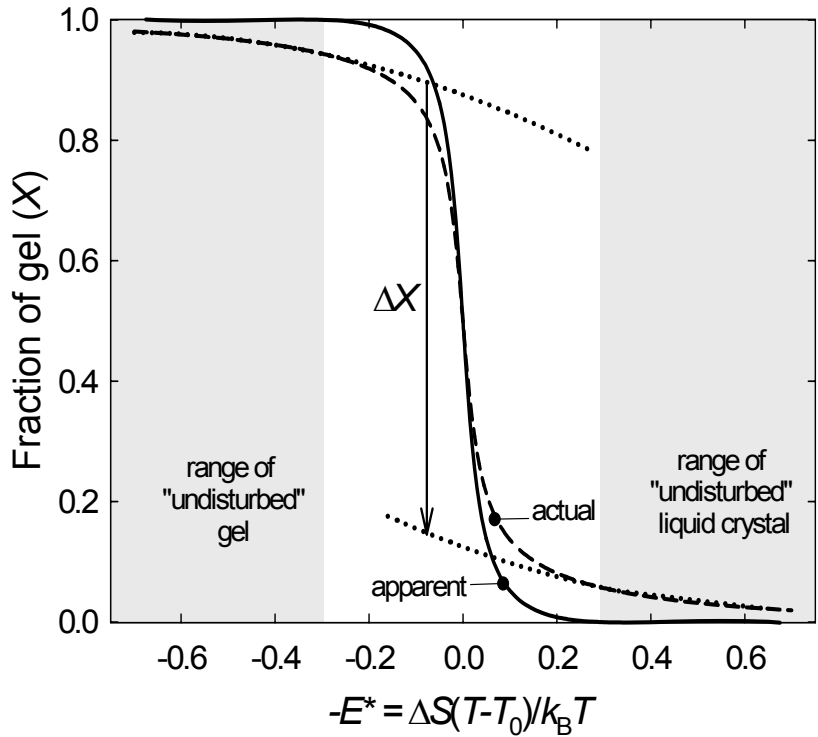

Fig. 2. A typical curve for the theoretical actual mole fraction of lattice points in gel-state, $X$, as function of reduced temperature $-\Delta E^*$ (dashed line), second polynomial approximation of "undisturbed" gel and liquid-crystalline states at $|\Delta E^*|>\Delta E^*_{lim}$ (dotted lines), and the apparent fraction of gel-state points in the lattice calculated according to Eq. 7. The ranges of "undisturbed" states are indicated in gray halftones.

To answer this question we should recall that, as mentioned in Introduction, a two-dimensional lattice model may apply to the phenomena that occur some distance away from the transition point, where the fraction of lipid embedded into the nuclei of the opposite state is low and the steric stress is negligible. Keeping this point in mind, we consider only the "tail" of the experimental temperature dependence and compare its shape with the theoretical curves. The experimental data for two lipids (DMPC measured in this work and DPPC studied earlier [17]) and the theoretical curves are presented in Fig.3 in reduced coordinates, $-\Delta E^*$ vs. $X_{app}$ (note that $-\Delta E^*$ is equivalent to the reduced temperature; see Model). Inspecting the figure, one reveals that the experimental curves correspond to $\varepsilon = 0.495$ for DMPC and $\varepsilon = 0.53$ for DPPC; both are below the critical value $\varepsilon = 0.55$ (see Table 1). It means the *g-l* contact energy in DMPC and DPPC is not high enough to cause discontinuity. In other words, *lateral interactions within monolayer cannot determine the first-order character of g-l transition*. One can therefore conclude that signs of the first-order behaviour observed in multilayer vesicles are due to the interlayer, three-dimensional interactions in vesicular systems.

The parameter $\varepsilon$ in DPPC is, however, closer to the critical value than in DMPC. It is expected, therefore, that in longer lipids, this parameter may be large enough to cause discontinuity. What is the critical chain length, above which the transition in lipid layers can be intrinsically first-order?

Considering the best fit values of $\varepsilon$, one finds out that $\varepsilon$ is proportional to the number of C-atoms in the lipid chains. This gives grounds for a linear extrapolation of the contact energy. Doing that, one finds the critical length to be about 17 C-atoms. This result is more clearly illustrated in Fig. 4, presenting the chain-length dependence of the direct *g-l* transition temperature $T_0$, the main transition temperature $T_m$, and the critical temperature $T_c$ estimated from $\varepsilon$ with Eq. 4. The critical temperature grows faster than $T_m$ or $T_0$, the intersection point being at 17 C-atoms. Taking into account the uncertainty in $\varepsilon$, one obtains a range of 16.5-18 C-atoms for the critical length.

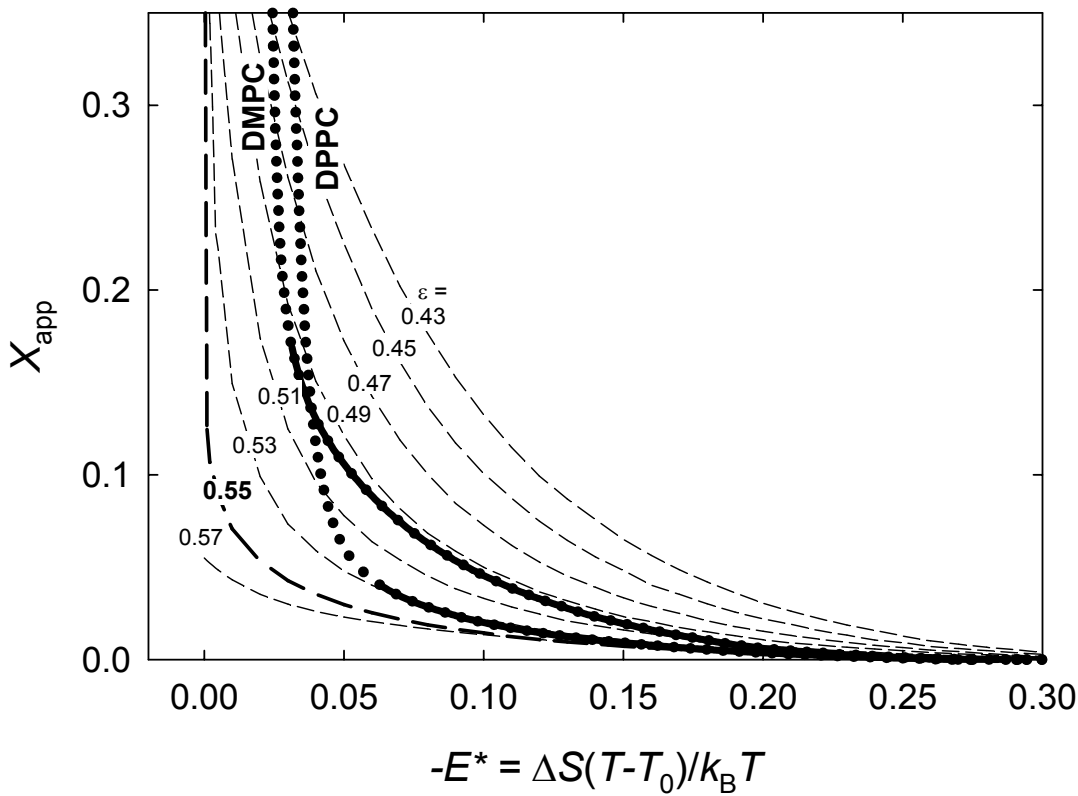

Fig. 3. Apparent fraction of gel as function of reduced temperature: a family of calculated curves at different $\varepsilon$ (dashed lines) and the experimental curves for DMPC and DPPC. The pretransitional ranges of experimental curves are shown in solid lines; and the ranges corresponding to the global chain-ordering transition of vesicles (where the experimental curves abruptly deviate from the theoretical ones) are shown in dotted lines.

Therefore, discontinuous chain-ordering transition should be inherent to monolayer when the lipid chains are at least 17 C-atoms or longer. In such long lipids, the first-order character of the transition is expected even in the systems where interactions in the third dimension are not significant. Other experimental data, supporting the singularity at this length, were reported in the recent work by Pabst et al. [18]. They found that the anomalous swelling in vicinity to the chain-ordering transition, a phenomenon caused by enhanced out-of-plane fluctuations, vanishes in the saturated PCs having more than 18 C-atoms per chain. This value is surprisingly close to that obtained in the present work. This suggests that the both pretransitional phenomena (the anomalous deviations of heat content and anomalous magnitude of out-of-plane motions) have the same origin: both are determined by lateral interactions.

The critical temperature obtained here should be considered in the spirit of the two-dimensional Ising model: long-range lateral order cannot exist at $T>T_c$ in the absence of additional (3D) interactions. Relation of this quantity to other definitions of critical temperature (e.g., that defined from the "lateral pressure/surface area" isotherms in lipid monolayers) is not yet clear and should be a matter of further study.

Therefore, we have found a critical chain length of 17-18 C-atoms for saturated PCs where the melting transition temperature equals to the critical temperature. This result may have an interesting biological implication. The saturated chains of this length are the most abundant in biological membranes. It could mean that biological membranes (or at least their regions enriched with saturated PCs) maintain themselves at a certain thickness corresponding to the critical state, in which the system is the most responsive to external physical actions. This property appears to be an important physical condition for functioning of the cellular membranes. This consideration is in line with the assertion that the chain-ordering/melting phase transition plays a crucial role in the membrane-related physiological mechanisms [1, 3, 28, 29].

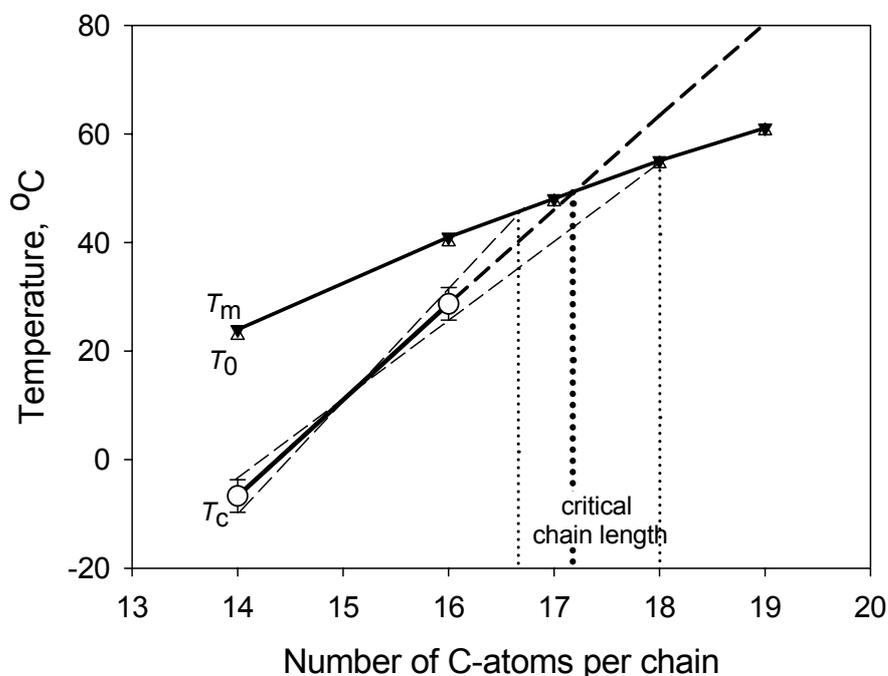

Fig. 4. Main transition ($T_m$), direct g-l transition ($T_0$), and critical ($T_c$) temperatures as functions of chain length. The values of $T_0$ for DMPC and DPPC are taken from Table 1; for longer lipids they have been estimated from data on pre- and main transition enthalpies ($H_{pre}$, $H_m$) and temperatures ($T_{pre}$, $T_m$), respectively (ref. [4], p.242), by means of equation $T_0(H_{pre}/T_{pre} + H_m/T_m) = (H_{pre} + H_m)$. Extrapolation of critical temperatures to longer lipids is shown with dashed line. The point of intersection of this line with the curves for $T_m$ ($\approx T_0$) corresponds to critical chain length; the maximum uncertainty interval of this parameter is indicated with thin vertical dotted lines.

## Acknowledgement

We thank Sergei Prosandeev, Sergei Pikin, Georg Pabst, Michael Rappolt, Peter Laggner, and Alexey Agafonov for fruitful discussions. The work was supported, in part, by INTAS (01-0105), RFBR (05-04-49206) and the Program of RAS for Fundamental Research in Molecular and Cell Biology.